\newif\ifSC
\SCtrue
\ifSC
\documentclass[onecolumn,draftclsnofoot,12pt]{IEEEtran}
\else
\documentclass[conference]{IEEEtran}
\fi

\usepackage{bbm}
\usepackage{color}
\usepackage{amsthm}

\usepackage{cite}
\usepackage{amsmath,amssymb,amsfonts}
\usepackage{algorithmic}

\usepackage{textcomp}
\usepackage{xcolor}
\newtheorem{theorem}{Theorem}

\newcommand{\ie}{{\em i.e. }}

\newcommand{\rd}{r_{\mathrm{d}}}

\newcommand{\sx}{s_{\mathbf{x}}}
\newcommand{\tbnproc}{\Phi}
\newcommand{\phim}{\Phi_\mathrm{M}}
\newcommand{\phitotal}{\Phi_\mathrm{T}}
\newcommand{\x}{\mathbf{x}}
\newcommand{\xd}{\mathbf{x}_\mathrm{d}}
\newcommand{\xt}{\mathbf{x}_\mathrm{t}}
\newcommand{\1}{\mathbbm{1}}
\newcommand{\norm}[1]{\| #1\|}

\newcommand{\erfc}[1]{\mathrm{erfc}\left\{#1\right\}}
\newcommand{\exps}{\mathbb{E}}
\newcommand{\ts}{T_{\mathrm{S}}}
\newcommand{\Es}{\mathrm{E_S}}

\newcommand{\Em}{\mathrm{E_M}}
\newcommand{\Et}{\mathrm{E_T}}

\newcommand{\expS}[1]{\exp\left(#1\right)}
\newcommand{\expU}[1]{e^{#1}}
\newcommand{\prob}[1]{\mathbb{P}\left[#1\right]}
\newcommand{\expect}[1]{\mathbb{E}\left[#1\right]}
\newcommand{\dd}{\mathrm{d}}
\newcommand{\pzero}{P_{\mathrm{0}}}

\newcommand{\pone}{P_{\mathrm{1}}}
\newcommand{\tbl}[1]{\tau_{\mathrm{L}#1 }    }

\newcommand{\tbh}[1]{\tau_{\mathrm{H}#1 }    }
\newcommand{\dfracnrho}[2]{\frac{\partial^{#1} #2}{\partial \rho^{#1} }}

\renewcommand{\pzero}{p_0}
\renewcommand{\pone}{p_1}
\newcommand{\probe}{\mathsf{P}_\mathrm{e}}
\newcommand{\probeb}[1]{\mathsf{P}_\mathrm{e#1}}
\newcommand{\dist}[2]{g_{\mathsf{R}_\mathrm{d}}\left(#2\right)}

\newcommand{\opt}[1]{#1_{\mathrm{opt}}}
\newcommand{\cb}{{\mathsf{b}}}

\newcommand{\resultheading}[1]{\vspace{-.09in} \ \\ \textit{\textbf{#1}}:}

\newcommand{\noisiqfunc}{P}

\newcommand{\cir}[2]{f(#1,#2)}
\newcommand{\cirslot}[1]{h_{#1}}

\ifCLASSINFOpdf
   \usepackage[pdftex]{graphicx}
  \graphicspath{{figs/}}
 \DeclareGraphicsExtensions{.pdf}
\else
\fi

\newcommand{\subsectioninline}[1]{\textbf{#1}:\ }

\title{Analysis of Diffusion Based Molecular Communication System with Multiple Transmitters
}
\IEEEoverridecommandlockouts
\begin{document}

	
\author{
	Nithin V. Sabu and Abhishek K. Gupta
	\thanks{The authors are with the department of Electrical Engineering, Indian Institute of Technology Kanpur, Kanpur, India 208016. (Email: nithinvs@iitk.ac.in and gkrabhi@iitk.ac.in). This research was supported by the  Science and Engineering Research Board  (DST, India) under the grant SRG/2019/001459. A part of the paper will appear in \cite{nvsabhi}.}}

\maketitle

\begin{abstract}
Due to the limited capabilities of a single bio-nanomachine, complicated tasks can be performed only with the co-operation of multiple bio-nanomachines. In this work, we consider a diffusion-based molecular communication system with a transmitter bio-nanomachine (TBN) communicating with a fully-absorbing spherical receiver bio-nanomachine (RBN) in the presence of other TBNs. The bits transmitted by each of the TBNs are considered as random in each time slot and different for each TBNs in contrary to the past works in literature with deterministic bits, which are same to all TBNs. The TBNs are modeled using a marked Poisson point process (PPP) with the location of TBNs as points of PPP, and the transmit bits as marks. In this paper, we derive the expected number of molecules observed at the RBN and the bit error probability of the system. We validate our analysis using numerical results and provide various design insights about the system.

\end{abstract}


\section{Introduction}
 Molecular communication can enable bio-nanomachines (biological devices with nanoscale functional units) to communicate with each other by sending and receiving messenger molecules termed as \textit{information molecules} (IMs). 
The transmitter bio-nanomachine (TBN) first encodes the transmit message into IMs \cite{nakano2013molecular}. 
Then, the TBN emits IMs to the propagation medium, and the IMs propagate to the receiver bio-nanomachine (RBN). In molecular communication via diffusion (MCvD) systems, the propagation is due to diffusion via Brownian motion \cite{einstein1956investigations}. 
The receptors present on the surface of the RBNs bind the IMs, and RBNs do further processing to estimate the transmitted information. \par
The channel model for a three dimensional (3D) MCvD system with a point TBN and a fully-absorbing receiver (absorbs all the IMs hitting its surface) was derived in \cite{yilmaz2014three}. The IMs of the same type emitted from the interfering TBNs also propagates to the receiver to cause multi-transmitter interference (MTI) \cite{pierobon2012intersymbol}. 
In literature, the spatial distribution of bacterial colonies inside cheese was shown to follow the Poisson point process PPP \cite{Jeanson1493}.  Therefore, the location of the bio-nanomachines in the 3D space can be modeled using PPP.
The expected number of molecules absorbed at the passive and fully-absorbing spherical receivers, when the TBNs are distributed as PPP was derived in \cite{deng2017analyzing}. The authors also derived the probability of bit error for the same system.
The work \cite{dinc2017theoretical} derived the expected number of molecules received at a fully-absorbing receiver by considering both inter-symbol interference (ISI) and MTI when the number of interfering TBNs is constant.
The analytical expression for the total signal strength and the bit error probability at the partially absorbing spherical receiver when the interfering transmitters are distributed as a PPP was derived in \cite{dissanayake2018enhancing}.\par
The expected absorbed molecules and the probability of bit error was calculated in the past works \cite{deng2017analyzing,dissanayake2018enhancing,dissanayake2019interference} by considering all the PPP distributed TBNs are sending the same bit sequence. 
In an MCvD with multiple TBNs, each TBN can have individual transmit data that may be distributed according to an arbitrary probability distribution over information symbols and can be independent of other TBNs. For such a system in practical scenarios, it is essential to include the independence and randomness of transmit data across TBNs in the system model. This was not studied in past works, which is the focus of this work.\par
In this work, we consider a 3D MCvD system with multiple interfering TBNs and a fully-absorbing spherical RBN. The location of the associated TBN is assumed to be fixed, or the associated TBN is the nearest transmitter. The interfering TBNs are modeled as a marked PPP (MPP) with their location as points of PPP and the transmit bits as marks. The transmit bits at each TBN is assumed to be random and independent of transmit bits at other TBNs.  In this paper, we derive the expected number of IMs observed at the RBN and the probability of bit error of the system. We also discuss the relevance of accurately incorporating the randomness of the data to be transmitted.

\section{System model}
\begin{figure}
	\centering
	\includegraphics[width=0.5\linewidth]{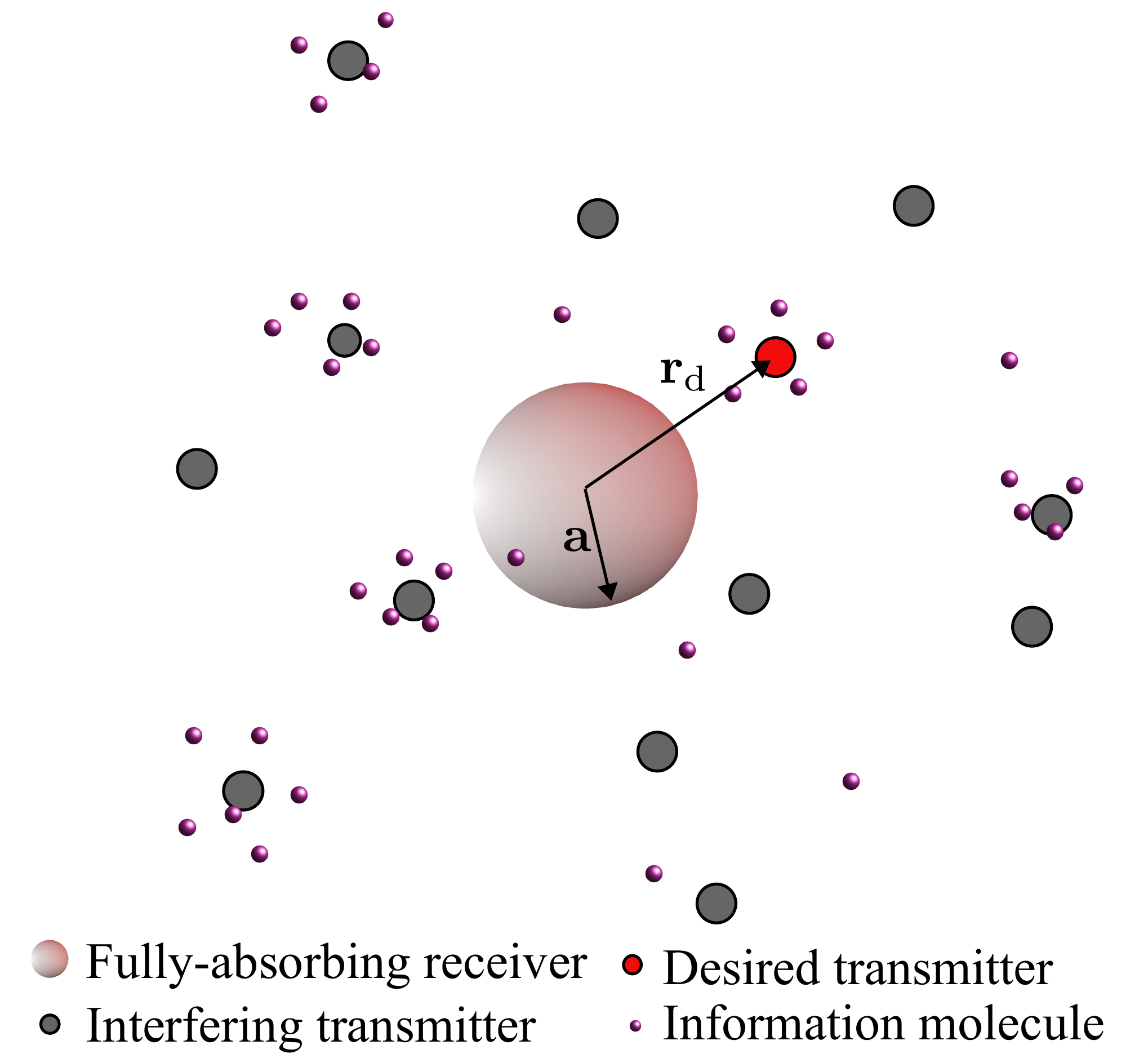}
	\caption{System model. An MCvD system with  a typical spherical fully absorbing RBN at the origin. The location of the desired TBN ($\xd$, shown as red circle) can either be fixed or nearest to the RBN and the interfering TBNs (shown as grey circle) form a MPP.}
	\label{sm}
\end{figure}
In this work, we consider an MCvD system in a 3D fluid medium without flow, with a fully absorbing spherical RBN and multiple TBNs, as shown in Fig. \ref{sm}. The TBNs are assumed to be point sources, which emit IMs to the fluid medium based on on-off keying (OOK) modulation.\par
\subsectioninline{Network Model}
Consider a typical fully-absorbing spherical RBN of radius $ a $ located at the origin. The whole surface of the RBN is covered with receptors that can bind only a single type of molecules. All the IMs reaching the surface of RBN are attached to the receptors and are counted for demodulation.\par
Assume that the TBN associated with the typical RBN, termed \textit{tagged} TBN, is located at $\xd$. We consider two cases; one with the tagged TBN at a fixed location and other with the nearest transmitter as the tagged TBN. The interfering TBN locations can be modeled using a 3D homogeneous PPP $\tbnproc$ \cite{dissanayake2019interference}. Since the RBN occupies the space $\mathcal{B}(0,a)$ (ball of radius $a$ centered at the origin), the support of PPP is taken as $\mathbb{R}^3\setminus\mathcal{B}(0,a)$ \cite{dissanayake2019interference}. The union of the location of the desired TBN and the interfering TBNs PPP ($\Phi$) is denoted by $\{\xd\}\cup \tbnproc$.

\subsectioninline{Modulation and transmission model}
Consider the transmit bit of the TBN located at $\x$ (denoted by $s_{\x}$), is assumed to be a Bernoulli RV with parameter $\pone$. At the beginning of the time slot (of duration $\ts$),  the point TBN emits $u_\x$ number of molecules into the fluid medium. The modulation scheme we consider is OOK. Therefore $u_\x=N$ when $\sx=1$ and $u_\x=0$ otherwise. $u_\x$ can be 0  with probability $\pzero=1-\pone$ and $N$ with probability $\pone$. $u_\x$ can be considered as the mark of the TBN, and the interfering TBNs can be modeled using a marked PPP $\phim=\{(\x,u_\x):\x\in\tbnproc\}$ \cite{Andrews2016a}. Also, the TBNs and RBN are assumed to be synchronized in time. 

\subsectioninline{Propagation model} Among various propagation mechanisms, we consider free diffusion for sending the IMs from TBN to RBN. The TBN emits the IMs to the propagation medium, and it moves to the RBN via 3D Brownian motion. 

\subsectioninline{Channel and receiving model}
Consider a TBN located at $\xt$ transmitting IMs to the propagation medium. The fraction of IMs observed at the RBN within time $t$ since the transmission at time $ t=0 $ be denoted by $\cir{t}{\norm{\xt}}$. The fraction of IMs observed at the RBN during the time interval $[0 ,\ts ]$ is
\begin{align}
\cirslot{\norm{\xt}}=\cir{\ts}{\norm{\xt}}\label{s341}.
\end{align}
$\cirslot{\norm{\xt}}$ is also known as channel impulse response (CIR).
Considering the event of observation of an IM  at the RBN  as a Bernoulli trial with probability of success $\cirslot{\norm{\xt}}$, the number of molecules $y_{\xt}$ observed at the RBN follows Binomial distribution with parameter $(N,\cirslot{\norm{\xt}})$, where $ N $ is the number of transmitted IMs. We can approximate Binomial distribution with Poisson distribution for mathematical tractability when $ N $ is large and $\cirslot{\norm{\xt}}$ is small. Hence $y_{\xt}\sim\mathcal{P}(N\cirslot{\norm{\xt}})$.
The total number of desired IMs reaching the RBN due to the emission of IMs from the tagged TBN is
\begin{align}
y_\mathrm{S}\sim\mathcal{P}\left( \cirslot{\norm{\xd}}u_{\xd}\right)\label{eq:yS}.
\end{align} 
 
Similarly, the total number of MTI molecules reaching the RBN due to the emission of IMs from the interfering TBNs, given $\tbnproc$, is
\begin{align}
y_\mathrm{M}&\sim \mathcal{P}\left( \sum_{\xt\in \tbnproc}\cirslot{\norm{\xt}}u_{\xt}\right)\label{eq:yM}.
\end{align}
Therefore, the total number of molecules absorbed by the RBN  at any time slot is the sum of the desired molecules and MTI molecules. 

\subsectioninline{Decoding at the RBN}
We consider a threshold detector at the RBN to demodulate the transmitted information. The RBN counts the total number of IMs absorbed ($y$) in a time slot, and at the end of the time slot, $ y $ is compared with a predefined threshold $\eta$. The bit $s_{\xd}$ transmitted from the tagged TBN is estimated as $\hat{s}_{\xd}=0$ if $y<\eta$, otherwise $\hat{s}_{\xd}=1$. Due to the diffusion of IMs and due to the presence of interfering TBNs in the propagation medium, errors can occur in the demodulation process. An error would occur at the RBN when the transmitted bit $s_{\xd}=0$ is decoded as $\hat{s}_{\xd}=1$ and vice versa. Therefore, the total probability of bit error ($\probe$) at any time slot is given by
\begin{align}
\probe= &\pzero \probeb{0}+\pone \probeb{1}\label{eq:berdef}
\end{align}
where $\probeb{0}$ and $\probeb{1}$ are the probability of incorrect decoding for bit 0 and 1 formally defined as $
\probeb{0}=
\prob{
	\hat{s}_{\xd}=1\mid s_{\xd}=0
} \text{ and } \
\probeb{1}=
\prob{
	\hat{s}_{\xd}=0\mid s_{\xd}=1
}.
$

\subsectioninline{Modeling molecular degradation}
The performance of a MCvD can be improved by incorporating adequate degradation of IMs in the design. IMs degrade over time due to the reaction with other molecules existing or added intentionally in the propagation medium.
We consider exponential degradation, where the probability that an IM will degrade only after time  $t$ is equal to $\exp{\left(-\mu t\right)}$. Here, $\mu$ denotes the reaction rate constant, and $\mu$ is related to the half-time ($\Lambda_{1/2}$) as $\mu=\ln(2)/\Lambda_{1/2}$.  $\mu\rightarrow 0$ ({\em i.e. } $\Lambda_{1/2}\rightarrow \infty$) corresponds to IM with no degradation.\par
We assume that the considered molecular communication system does not have inter symbol interference. Some examples of such systems include cases where the symbol time $\ts$ is sufficiently large and/or molecular degradation rate is sufficient. 

\subsectioninline{Channel impulse response}  The hitting rate of molecules  at the surface of the RBN (\ie total number of molecules hitting the RBN in unit time) at time $\tau$, due to the emission of IMs from a point TBN located $r$ distance away from the center of RBN is given as \cite{yilmaz2014three},
\begin{equation}
\kappa(\tau, r)=\frac{a}{r}\frac{r-a}{\sqrt{4\pi D \tau^3}}\exp\left( -\frac{(r-a)^2}{4D\tau}\right) ,\label{s31}
\end{equation}
where $D$ represents the diffusion coefficient, which depends on the properties of IM and the propagation medium.  The fraction of non-degraded IMs observed at the RBN within time $t$, is given by \cite{heren2015effect},
\begin{align}
&\cir{t}{r}=\int_{0}^{t}\kappa(\tau, r)\times \expS{-\mu \tau}\dd\tau\\
=& \frac{a}{2r}\left[\exp\left(-\sqrt{\frac{\mu}{D}}(r-a)
\right)\erfc{\frac{r-a}{\sqrt{4Dt}} -\sqrt{\mu t}}
+\exp\left(\sqrt{\frac{\mu}{D}}(r-a)
\right)\erfc{\frac{r-a}{\sqrt{4Dt}} +\sqrt{\mu t} }\right]\label{s32}.
\end{align}

\subsectioninline{Observations at the RBN}
The total number of IMs arriving at the RBN due to the emission of IMs from the tagged and interfering TBNs is $ y=y_\mathrm{S}+y_\mathrm{M}.$
Using \eqref{eq:yS}, \eqref{eq:yM}, and since the sum of Poisson random variables is also Poisson random variable, $ y\sim \mathcal{P}\left( \sum_{\xt\in \phitotal}h_{\norm{\xt}}u_{\xt}\right) $.
From \eqref{eq:yS}, the expected number of desired IMs observed at the RBN is
\begin{align}
\Es &=\expect{y_{\mathrm{S}}}=
\pone N\cir{\ts}{\rd}.\label{eq:averageES}
\end{align}
The expected number of molecules arriving at the RBN due to MTI is 
\begin{align}
	\Em =\exps\left[y_{\mathrm{M}}\right]&=\mathbb{E}\left[\sum_{\x\in \tbnproc}\cirslot{\norm{\x}}u_{\x}\right].\label{eq:averageEM}
	\end{align}
Applying Campbell Mecke theorem \cite{haenggi2012stochastic} in \eqref{eq:averageEM} gives,
\begin{align}
\Em &=4\pi\lambda \int_{a}^{\infty}h_{z}\mathbb{E}[u_{z}]z^2\mathrm{d}z\nonumber\\
&
=4\pi\lambda  \pone N\int_{a}^{\infty} \cir{\ts}{z} z^2 \mathrm{d}z\nonumber.
\end{align}
Now from \eqref{eq:averageES} and \eqref{eq:averageEM}, the expected total number of IMs absorbed by the RBN ($\Et =\Es  +\Em$) at any time slot is
\begin{align}
\Et =\pone N\left(\cir{\ts}{\rd}+4\pi\lambda \int_{a}^{\infty}
\cir{\ts}{z}
z^2
\dd z\right).\label{eq:averageET}
\end{align}
\textit{Special case: }When $\ts\rightarrow \infty$,
	\begin{align}
\Em &=4\pi\lambda \pone Na\left(\frac{D}{\mu}+a\sqrt{\frac{D}{\mu}}\right).\label{eq:averageEMTs}
\end{align}
In \eqref{eq:averageEMTs}, when $\ts\rightarrow \infty$, the expected number of observed molecules due to MTI increases with  $\lambda , \pone, N , a$ and $D$, and decreases with $\mu$. Also in \eqref{eq:averageEMTs}, we can see that when $\mu\rightarrow 0$ (no molecular degradation), $\Em \rightarrow \infty$. This implies that, for a system with IM does not degrade over time, the expected MTI molecules will tend to infinity when $\ts\rightarrow \infty$. \par

\section{Probability of bit error}
The probability of bit error as defined in \eqref{eq:berdef} for the considered MCvD system is derived in this section. First, we consider the distance $\rd$ between the tagged TBN and the RBN is fixed, and we derive the $\probe$. We then obtain $\probe$ when the desired transmitter is the nearest TBN.

\subsection{When the tagged TBN is at a fixed distance:}
The probability of bit error for the case when $\rd$ is fixed is given in Theorem \ref{thm:fixedrdnoisi}.
\begin{theorem}\label{thm:fixedrdnoisi}
When the tagged TBN is at a fixed distance from the RBN, the probability of bit error is given by \eqref{eq:berdef} with the probability of incorrect decoding of bit 0 and 1 given as

\begin{align}
\probeb{0}&=1-\expU{-\alpha_0(\lambda)} 
\left[ 			
1+\sum_{n=1}^{\eta-1}\frac{1}{n!}\mathfrak{B}_n (\pmb{\alpha}(\lambda))
\right],
\label{eq12}\\
\probeb{1}&=
\expU{-\alpha_0(\lambda)-  N\cir{\ts}{\rd}} 
\times \left[1+\sum_{n=1}^{\eta-1}\frac{1}{n!}\mathfrak{B}_n (\pmb{\beta}(\rd,\lambda))\right],\label{eq13}
\end{align}

where
$
\alpha_0(\lambda)=4\pi \lambda \pone \int_{a}^{\infty}
	\left[
	1-\expU{ - N\cir{\ts}{z}}
	\right]
	z^2\dd z,
$ $\pmb{\alpha}(\lambda)=[\alpha_{1}(\lambda),\alpha_{2}(\lambda),...,\alpha_{\eta-1}(\lambda)]$ 
and $\pmb{\beta}(\rd,\lambda)=[\alpha_{1}(\lambda)+N\cir{\ts}{\rd},\alpha_{2}(\lambda),...,\alpha_{\eta-1}(\lambda)]$ with
\begin{align}
&\alpha_i(\lambda)= 4\pi \lambda \pone \int_{a}^{\infty}
\expU{
	-N\cir{\ts}{z}
}
 {\left(N\cir{\ts}{z}\right)}^i 
 z^2\dd z.
\end{align}
Here, $\mathfrak{B}_n(.)$ denotes the $n$th complete exponential Bell polynomial \cite{rio} given as
\begin{align}
&\mathfrak{B}_{n} (\pmb{\alpha}(\lambda))=\sum_{w=1}^n\sum\frac{n!}{j_1!j_2!...j_{n-w+1}!} \prod_{v=1}^{n-w+1}\left(\frac{\alpha_v(\lambda)}{v!}\right)^{j_v}.\label{pbp}
\end{align}
where the second sum is taken over all non-negative integers $j_1,j_2,...,j_{n-w+1}$ such that $j_1+j_2+...+j_{n-w+1}=w$ and $1j_1+2j_2+...+(n-w+1)j_{n-w+1}=n$.
\end{theorem}
\begin{IEEEproof}
	See Appendix \ref{app:C}.
\end{IEEEproof}

\subsection{When the tagged TBN is the nearest transmitter:}
Now, consider the case when the nearest transmitter is the tagged TBN. This case is more realistic as the tagged TBN location is not fixed. The probability density function of $\rd$ is 
\begin{align}
\dist{\rd}{r}=4\pi \lambda r^2\exp\left(-\frac{4}{3}\pi\lambda\left(r^3-a^3\right)\right) 
\end{align}
The probability of bit error for this case is given in Theorem \ref{thm:randomrdnoisi}. The proof is very similar to the proof of Theorem 1 and hence is omitted for brevity.
\begin{theorem}\label{thm:randomrdnoisi}
	When the nearest TBN is selected as the tagged TBN, 
	 the probability of bit error is given by \eqref{eq:berdef} with the probability of incorrect decoding for bit 0 and 1 given as 
		\begin{align}
	\probeb{0}=&1-\expU{-\alpha_0(\lambda)} 
	\left[ 			
	1+\sum_{n=1}^{\eta-1}\frac{1}{n!}\mathfrak{B}_n (\pmb{\alpha}(\lambda))
	\right],
	\label{eq20}\\
	\probeb{1}=&	
	4\pi \lambda\expU{-\alpha_0(\lambda)+\frac{4}{3}\pi \lambda a^3} 
	\int_a^\infty\left[1+\sum_{n=1}^{\eta-1}\frac{1}{n!}\mathfrak{B}_n (\pmb{\beta}(\rd,\lambda))\right]
	\nonumber\\
	&
	\times\expS{-  N\cir{\ts}{\rd}-\frac{4}{3}\pi\lambda\rd^3
	}\rd^2\dd \rd
	\label{eqe19}
	\end{align}
where  $\alpha_0(\lambda),\ \pmb{\alpha}(\lambda)$ and $\pmb{\beta}(\rd,\lambda)$ are the same as in Theorem \ref{thm:fixedrdnoisi}.
\end{theorem}
\section{Numerical Results}
In this section, the analytical expressions derived in the previous sections are validated using Monte Carlo based simulations, and several design insights about the system are discussed with the help of numerical results. \par
For the Monte Carlo simulation, the interfering TBNs are generated as PPP outside the RBN and up to a distance of $150\mu m$ from the center of the RBN.  The simulation is carried out for $10^4$ realizations of PPP. The interfering TBN density chosen for simulations is $10^{-5}$ TBNs per $\mu m^3$, which corresponds to 141 interfering transmitters. For all simulations, the diffusion coefficient is fixed as $D= 74.9 \ \mu m^2/s$, fully-absorbing spherical receiver radius is fixed as $a= 4\mu m$, the number of molecules emitted for bit-$1$ is $N=100$ molecules, degradation rate constant $\mu=5s^{-1}$, and the duration of the time slot is set as $0.5$s. The chosen value of $\mu$ and $\ts$ ensure that the ISI is negligible. In all figures, solid lines represents the curves corresponding to the derived analytical expressions, and markers represent the simulation results unless otherwise mentioned.\par

\resultheading{Variation of $\Es,\ \Em$ and $\Et$ with  the distance between the tagged TBN  and the RBN $(\rd)$}
\begin{figure}
	\centering
	\includegraphics[width=.55\linewidth]{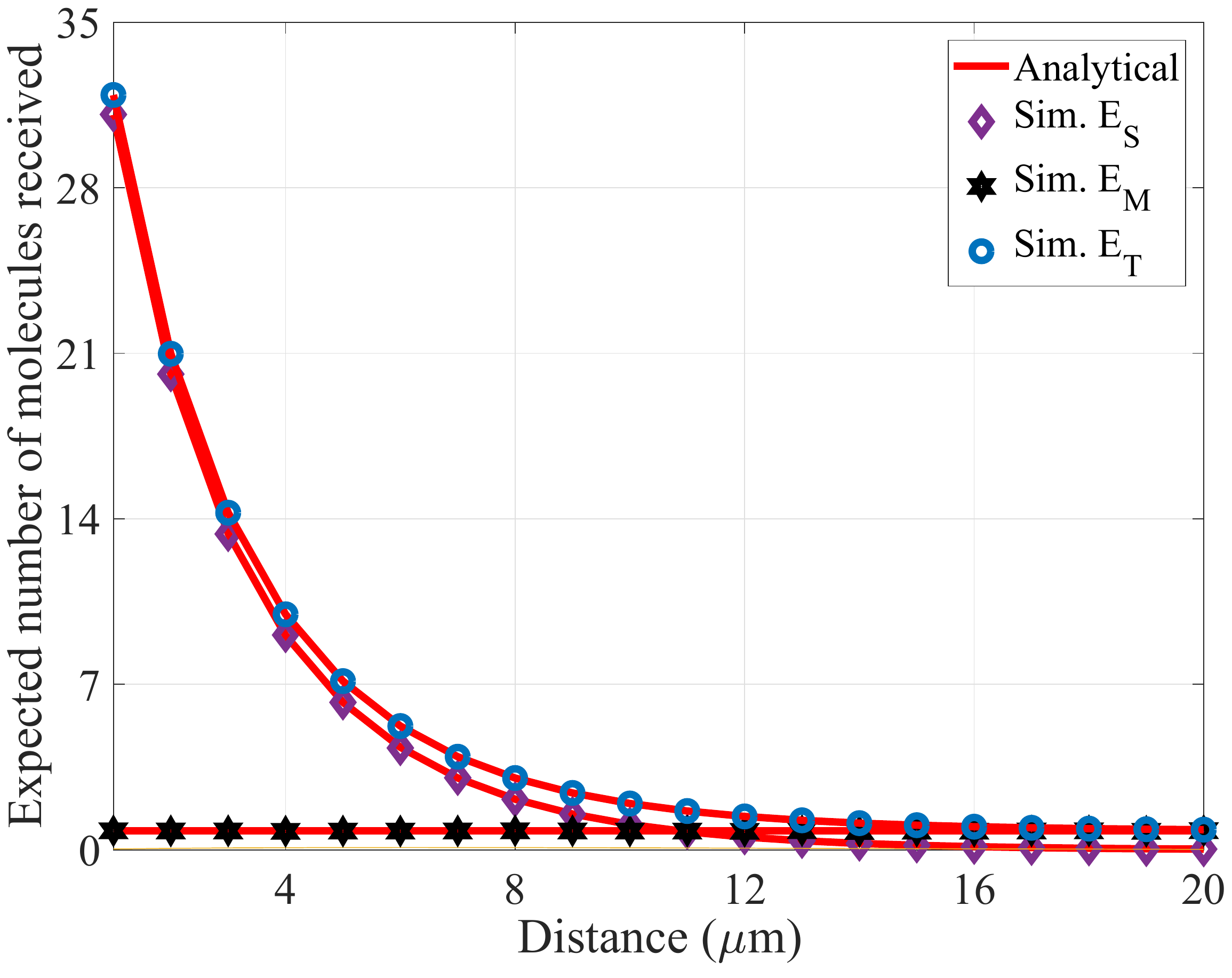}
	\vspace{-0.3cm}
	\caption{Variation of the expected number of desired ($\Es $), MTI ($\Em $) and total ($\Et $) molecules absorbed at the receiver for a system with molecular degradation versus the distance between tagged TBN and the RBN's surface $(\rd-a)$. Here, $ \lambda = 1\times 10^{-5} \text{ TBNs}/ \mu m^3$.}
	\label{fig:f4}
	\vspace{-0.3cm}
\end{figure}
The variation of $\Es,\ \Em$, and $\Et$ with the distance between the surface of the spherical RBN and the tagged TBN ( \ie, $\rd-a$) can be seen in Fig. \ref{fig:f4}. The observation of the expected number of absorbed MTI molecules ($\Em$) at the RBN is independent of $\rd$. The tagged TBN location affects the number of desired IMs absorbed at the RBN ($\Es$). $\Es$ and $\Et$ decreases as the tagged TBN moves away from the RBN. When the tagged TBN moves far away from the RBN, $\Es$ reduces to zero, and $\Et$ is only due to $\Em$, which results in bit error and loss of information.
Since $\Et$ varies with $\rd$, the decoding threshold $\eta$ at the RBN should be chosen according to  $\rd$.

\resultheading{Impact of decoding threshold $(\eta)$ on the probability of bit error $(\probe)$  when the location of tagged TBN is fixed} 
\begin{figure}
	\centering
	\includegraphics[width=0.55\linewidth]{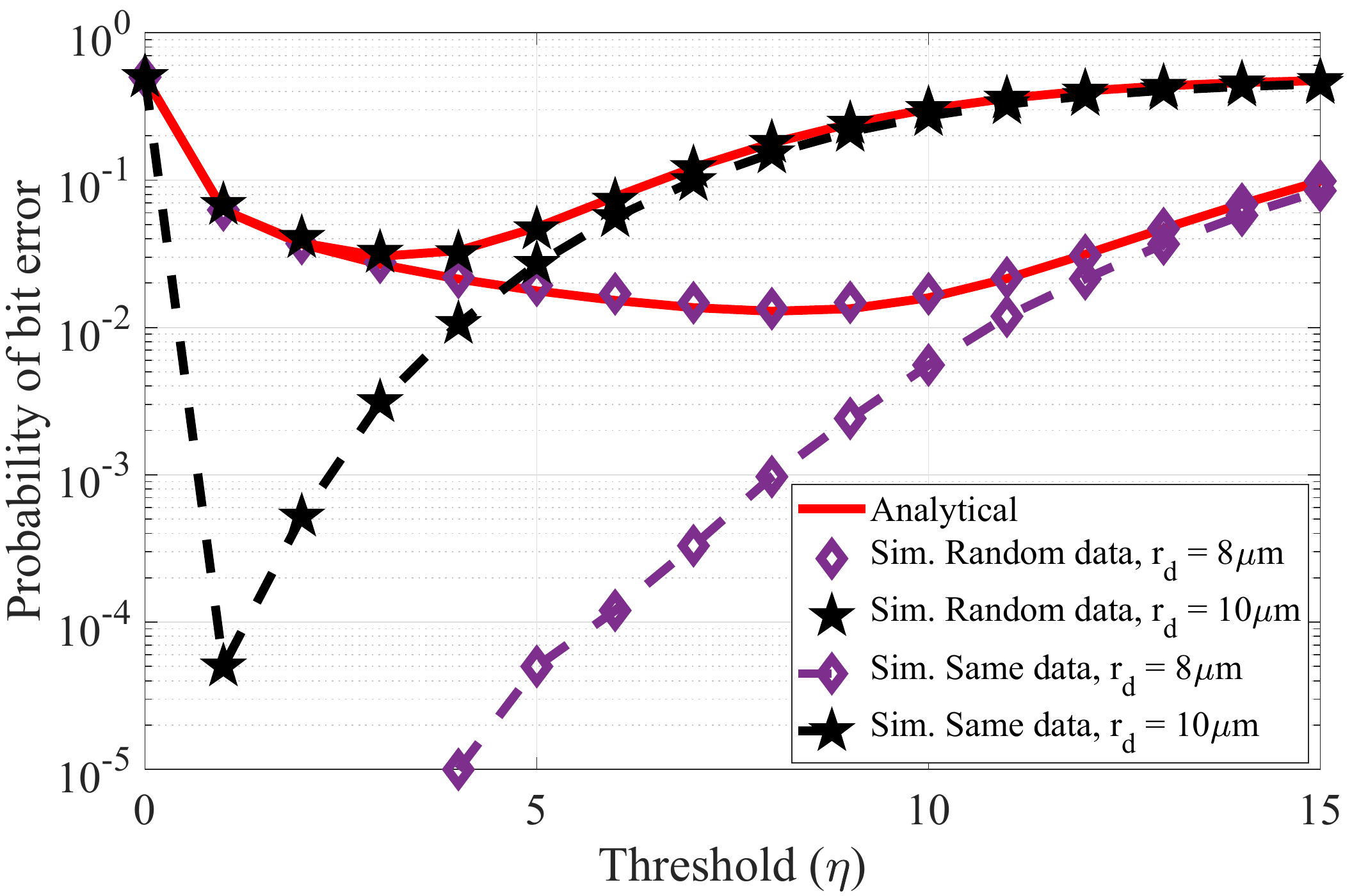}
	\caption{Probability of bit error ($\probe$) versus threshold ($\eta$) when $\rd$ is fixed. When $\rd$ increases, $\probe$ increases while the optimal threshold $\opt{\eta}$ decreases.}
	\label{fig:f5}
\end{figure}
Fig. \ref{fig:f5} shows the variation of $\probe$ with the threshold $\eta$ when the tagged TBN location is fixed.
Analytical results when transmit bits are random and different for each TBN (as derived in Theorem \ref{thm:fixedrdnoisi}) are compared with corresponding simulation results for various values of $\rd$. As the threshold for detection increases, the probability of bit error $\probe$ increases after it is first reduced to a minimum value. This behavior proves the existence of an optimum threshold $\opt{\eta}$ for which $\probe$ is minimum. When $\rd$ increases, $\Et$ reduces due to the reduction in $\Es$, and as a result, $\opt{\eta}$ decreases. The probability of bit error at the optimum threshold increases with $\rd$ due to the relative reduction in $\Es$ in comparison to $\Em$.\\
\resultheading{Impact of accurately characterizing randomness and independence of transmit data across TBNs}
In Fig. \ref{fig:f5} we can observe that, considering transmit bits same for all TBNs as in previous works give inaccurate results (especially at low $\eta$ values) compared to real scenarios (bits are random and different for TBNs).


\resultheading{Impact of decoding threshold $(\eta)$ on the probability of bit error $(\probe)$  when the tagged TBN is the nearest TBN}
\begin{figure}
	\centering
	\includegraphics[width=0.55\linewidth]{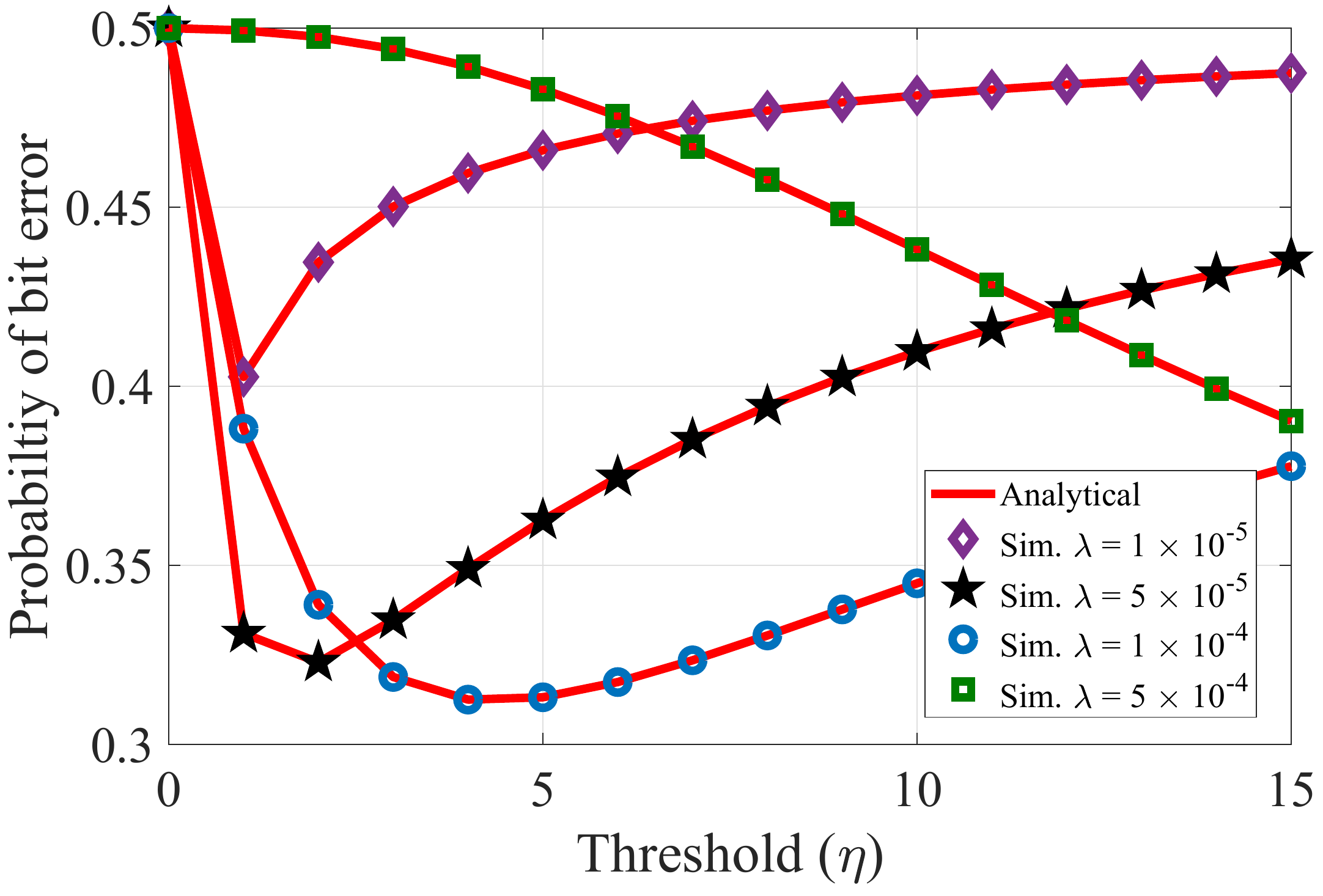}
	\caption{Probability of bit error versus the threshold $\eta$ with different interfering transmitter densities. Here, the tagged TBN is the nearest transmitter.}
	\label{fig:wcncf3}
\end{figure}
Fig. \ref{fig:wcncf3} shows the variation of $\probe$ with the detection threshold for different TBN densities. Similar to Fig. \ref{fig:f5}, as $\eta$ increases, $\probe$ first decreases and achieves a minimum value and after that, $\probe$ increases.  We can observe that, when TBN density increases, the minimum probability of bit error reduces.  This is because the nearest TBN is the desired transmitter, and when the TBN density increases, the desired transmitter comes closer to the receiver, and more signal molecules reach the receiver. Also, with the increase in TBN density, the optimal threshold $\opt{\eta}$ increases due to more signal molecules reaching the receiver.
\section{Conclusions}
In this paper, we have presented an analytical framework for a 3D MCvD system with multiple point TBNs and a single fully-absorbing spherical RBN. The data transmitted by each TBNs are random independent of other TBNs. The analytical expressions for the expected number of signal and MTI molecules absorbed by the RBN were derived. We have also derived the analytical expressions for the probability of bit error under two scenarios; when the desired TBN is at a fixed location, and the desired TBN is the nearest transmitter. The necessity of incorporating the transmission data randomness and independence is also included in the discussion. As future work, we can consider the impact of leftover molecules from previous symbols (which can occur when the symbol duration is small and/or degradation of IMs is not adequate) along with MTI on the system performance.

\appendices

\section{Proof of Theorem-1}\label{app:C}
Consider the bit to be transmitted by the tagged TBN  in a time slot as $\cb$.  Hence, $u_{\xd}=\cb N$. 

Let $v(\norm{\x})$ be the expected number of IMs absorbed by the RBN, that were emitted by the transmitter located  at $\x$ {\em i.e.}, $v(\norm{\x})=\cirslot{\norm{\x}} u_{\x}$.
Let $V$ be  the expected total number of received molecules conditioned on $\phim$ {\em i.e.},
$
V(\rd,\cb,\phim)=\cb N\cirslot{\rd}+\sum_{\x\in \phim} v(\norm{\x}).
$
Given $\phim$, the total number of IMs absorbed at the TBN is Poisson distributed \ie 
$
y\mid \phim \sim \mathcal{P}(V(\rd,\cb,\phim)).\label{eq:AppCymdist}
$
Therefore, the probability of incorrect decoding for bit $\cb$ is 
$\probeb{\cb}=\prob{y\notin [\tbl{\cb}\ \tbh{\cb} ]}$,
where $\tbh{\cb}$ and $\tbl{\cb}$ are upper and lower threshold value of bit $\cb$. Here, $\tbh{0}=\eta-1$, $\tbl{0}=0$, $\tbh{1}=\infty$, and $\tbl{1}=\eta$.  Now, 
the probability of incorrect decoding for bit $\cb$ is given as
\begin{align}
\probeb{\cb}&=
1-\sum\nolimits_{n=\tbl{\cb}}^{\tbh{\cb}} 
\mathbb{E}_{\phim}\left[
\prob{y=n\mid \phim}
\right] \nonumber\\
&=1-\sum\nolimits_{n=\tbl{\cb}}^{\tbh{\cb}} 
\mathbb{E}_{\phim}\left[
\frac1{n!}\exp\left(-V(\rd,\cb,\phim)\right)\times V(\rd,\cb,\phim)^n
\right] .\label{eq:AppC1}
\end{align}

\newcommand{\laplace}[2]{\mathcal{L}_{#1}\left(  #2\right)}
Note that, 
$
\expect{Z^n\expU{-Z}}=
{(-1)}^n
\left.
\frac{	
	\dd^n \laplace{Z}{\rho}
}
{
	\dd\rho^n
}
\right\vert_{\rho=1}.
$

Applying this identity in \eqref{eq:AppC1}, we get
\begin{align}
\probeb{\cb}
&=1-\sum\nolimits_{n=\tbl{\cb}}^{\tbh{\cb}} 
\frac1{n!}{(-1)}^n
\left.
\frac{	
	\dd^n \laplace{V(\rd,\cb,\phim)}{\rho}
}
{
	\dd\rho^n
}
\right\vert_{\rho=1},
\label{eq31}
\end{align}
with the slight abuse of notation that $\dfracnrho{n}{F}=F$ for $n=0$. 
In \eqref{eq31}, $ \mathcal{L}_{V}(\rho)$ is the Laplace transform of $V$ which can be obtained as,
\begin{align}
&\laplace{V(\rd,\cb,\phim)}{\rho}=\mathbb{E}\left [ \exp\left( -\rho\cb N \cirslot{\rd}
-\rho\sum\nolimits_{\x\in \phim} v(\norm{\x}) 
\right)\right ]\nonumber\\
&=\exp\left(-\rho \cb N h_{\rd}\right)
\mathbb{E}_{\phim}
\left [
\exp\left( -\rho\sum\nolimits_{\x\in  \phim } v(\|\x\|) \right)\right ]\nonumber\\
&\stackrel{(a)}{=}
\exp\left (
-\rho \cb N h_{\rd}-4\pi \lambda\int_{a}^{\infty}\left(1-\mathbb{E}_{u_{z}}
\left[
\expU{	-\rho h_{z}u_{z}}
\right]\right)
\ z^2\mathrm{d}z
\right )\nonumber\\
&=\exp\left (-
\rho N\cb h_{\rd}-4\pi \lambda \pone  \int_{a}^{\infty}
\left(
1-\expU{-\rho h_{z}N}
\right)\ \ z^2\dd z\right ),\label{gpe4}
\end{align}
where $(a)$ is due to the marked version of Campbell theorem.
By taking the $ n $th derivative of  \eqref{gpe4} using the Bell polynomial version of Faa di Bruno's formula \cite[eq.(2.2)]{rio}, we get
\begin{align}
\left.\frac{\dd^n \mathcal{L}_{V}(\rho)}{\dd\rho^n}\right\vert_{\rho=1}&=
\mathfrak{B}_{n} (\mathbf{\noisiqfunc}(\rd,\lambda))\times { (-1)}^n
\expS{ - \rho\cb Nh_{\rd}-4\pi \lambda 
	\pone
	\int_{a}^{\infty}(1 -\exp( - h_{z}N ))\ z^2\mathrm{d}z}
,\label{eq33}
\end{align}
 where $\mathbf{\noisiqfunc}(\rd,\lambda){=}[\noisiqfunc_{1}(\rd,\lambda),\noisiqfunc_{2}(\rd,\lambda),\cdots,\noisiqfunc_{\eta-1}(\rd,\lambda)]$ with $
\noisiqfunc_m(\rd,\lambda)= N\cb \cirslot{\rd}
\1(m=1)+4\pi \lambda 
\pone \int_{a}^{\infty}\expU{ -h_{z}N} {(h_{z}N )}^m z^2\dd z$. Now, substitute \eqref{eq33} in \eqref{eq31} with $\cb=0$ and $\cb=1$, we get \eqref{eq12} and \eqref{eq13} respectively.
\ifCLASSOPTIONcaptionsoff
  \newpage
\fi

\bibliographystyle{IEEEtran}
\bibliography{References}

\end{document}